\documentclass{ws-procs9x6}

\newcommand{\lsim}{\raisebox{-0.13cm}{~\shortstack{$<$ \\[-0.07cm] $\sim$}}~}

\begin{document}
\title{MSSM HIGGS BOSON PRODUCTION FROM SUPERSYMMETRIC CASCADES AT LHC
\footnote{\uppercase{T}alk presented by \uppercase{A.K.D.}
at {\it \uppercase{SUSY} 2003:
\uppercase{S}upersymmetry in the \uppercase{D}esert}\/, 
held at the \uppercase{U}niversity of \uppercase{A}rizona,
\uppercase{T}ucson, \uppercase{AZ}, \uppercase{J}une 5-10, 2003.
\uppercase{T}o appear in the \uppercase{P}roceedings.}
}
%%  PLEASE USE THE ABOVE FOOTNOTE FOR ALL ARXIV POSTINGS,
%%   (SUBSTITUTING SPEAKER NAME, OF COURSE;  NO SPEAKER NAME NEEDED
%%      FOR SINGLE-AUTHOR CONTRIBUTIONS). 
%%  NOT NECESSARY TO HAVE FOOTNOTE FOR VERSIONS THAT ARE SENT TO 
%%  SUSY 2003 CONFERENCE FOR PUBLICATION

\author{ASESH K. DATTA}

\address{Institute for Fundamental Theory,
Department of Physics \\
University of Florida, 
Gainesville, FL 32611, USA\\ 
%E-mail: datta@phys.ufl.edu
}


\maketitle

\vspace{-0.3cm}

\abstracts{
Detectability of MSSM Higgs bosons in cascade decays of supersymmetric 
particles at the CERN LHC is discussed. }
\vspace{-0.8cm}

\section{Introduction}
The Minimal Supersymmetric Standard Model (MSSM) have got 5 
physical Higgs scalars: 2 neutral CP-even, $h$ (the lightest one)
and $H$; 1 neutral CP-odd, $A$; and 
2 charged ones, $H^\pm$. The search strategies for these Higgs bosons
have thus far been based mainly on their direct
productions through Standard Model (SM)--like processes \emph{viz.}, 
$gg \to h,H,A$, $gg/ q\bar{q} \to h,H,A+ b\bar{b} / t\bar{t}$
for the neutral Higgses and $t \to H^+ b$, 
$gg/ q\bar{q} \to H^+ b\bar{t}$ and $gb
\to H^-t$ for the charged Higgs (see Ref.~\cite{reviews,TDRs} 
and references therein). 
Rates of most of these processes are strongly
enhanced as $\tan \beta$ grows. 
In this talk I briefly explore another potential source of MSSM Higgs bosons 
at the CERN Large Hadron Collider (LHC) in cascade decays of squarks 
and gluinos.

\section{The scheme}

If kinematically allowed, squarks and gluinos, copiously produced at 
LHC, could undergo 
the following cascade patterns in their decay:
\begin{eqnarray}
\hspace*{-0.4cm}
pp \to \tilde{g} \tilde{g} , \tilde{q} \tilde{q}, \tilde{q} \tilde{q}^* ,
\tilde{q} \tilde{g}  & \to & \chi_2^\pm  , \chi_3^0 , \chi_4^0 + X  
% \nonumber \\
%& \to &   
\to \chi_1^\pm , \chi_2^0 , \chi_1^0  + h,H,A, H^\pm + X \\
%\end{eqnarray}
%\begin{eqnarray}
\hspace*{-0.4cm}
pp \to \tilde{g} \tilde{g} , \tilde{q} \tilde{q}, \tilde{q} \tilde{q}^* ,
\tilde{q} \tilde{g}  & \to & \chi_1^\pm, \chi_2^0 + X  
%\nonumber \\
%&\to&
\to    \chi_1^0 + H^\pm, h,H,A \ +X
\end{eqnarray}
We call the decay chain in eq.~(1) the ``big cascade" while
the one in eq.~(2) is dubbed the ``little cascade''.
Other possibilities include:
\begin{eqnarray}
\hspace*{-0.4cm}
pp &\to & \tilde{t}_2 \tilde{t}_2^*, \tilde{b}_2 \tilde{b}_2^* \ \ {\rm with} \ \
\tilde{t}_2 (\tilde{b}_2) \to \tilde{t}_1 (\tilde{b}_1) +h/H/A \ {\rm or} \
\tilde{b}_1 (\tilde{t}_1) +H^\pm \\
%\end{eqnarray}
%\begin{eqnarray}
\hspace*{-0.4cm}
pp & \to &  \tilde{g} \tilde{g} , \tilde{q} \tilde{q}, \tilde{q} \tilde{q}^* ,
\tilde{q} \tilde{g}   \to  t \bar{t} + X  \to    H^\pm +X
\end{eqnarray}
Little cascades were discussed earlier \cite{hcascade,oldcascade}
for $h$ and relatively light $A,H$ and $H^\pm$. We reanalyze this case
in a broader perspective along 
with the newly proposed big cascades \cite{H+cascade}. 
Fast simulations for the signal and the backgrounds are 
performed including CMS detector response at the LHC.

\section{The motivation}
The motivating factors are: 
(i) couplings involved in cascades are ingredients of weak scale SUSY 
Lagrangian and would bear informations on the electroweak symmetry breaking
sector;
(ii) existence of a hole (where only the lightest $h$ boson can be found at 
the LHC) in the canonical reach plot 
spanning over $130 \lsim M_A \lsim 170$ GeV and 
$\tan\beta \sim 5$ is traced 
back to insufficient production rates at lower $\tan\beta$. 
With SUSY cascades as the dominant source of Higgses, 
$\tan\beta$ dependence gets diluted. This could fill up the hole;
(iii) SUSY cascades are sources
of MSSM Higgs bosons of much unforeseen potential and hence, on its own
right, must be analyzed anyway.

\section{Analysis and Observations}
We recently made \cite{H+cascade} a detailed analysis over the allowed 
parameter space. However, in this talk I only discuss to what extent
can one fill up the above-mentioned hole and the possible reach for
different Higgses thereon.

%\subsection{The Framework}

We make no assumptions to relate the two Higgs mass parameters, 
$m_{H_1}$ and
$m_{H_2}$, and the masses of squarks and/or sleptons. 
However, gaugino mass unification at a high scale is assumed. 
A somewhat large value of the trilinear $A$ parameter ($A_t$=$A_B$=1.5 TeV)
helps evade the LEP bound on $m_h$ and leads to the typical mixing scenario. 
The Higgs sector is treated with {\tt HDECAY} and CTEQ3L \cite{hdkcteq}
parton distributions are used. To be conservative,
$K$--factors and QCD corrections to squark/gluino decays 
are not considered.

%\subsection{Signal rates and cascade suppressions}

The study exploits the large rates for squarks and
gluinos at LHC. This is a prerequisite for healthy Higgs rates 
under cascades which typically involve an effective
branching ratio of only a few percent  for \emph{single} Higgs final 
states. 
Typical total rate for squarks and gluinos is 
$\sim 110(3)$ pb for $m_{\tilde{g}} \sim m_{\tilde{q}} \sim 0.5(1)$  TeV. 
This leads to a large 
($\sim 10^{6}(10^{5})$) number of parent events for the cascades
with an  accumulated luminosity $(\int {\mathcal L}) \sim 30$ fb$^{-1}$.

A key point here is that the couplings of the Higgs bosons to
charginos and neutralinos are maximal for higgsino-gaugino mixed states
\cite{Haber}. This results in dominant decays of the heavier chargino and
neutralinos into the lighter ones and Higgs bosons. A similar argument holds 
for the little cascades in the gaugino region whenever kinematically allowed.

For illustration I consider a generic scenario with 
$m_{\tilde{g}}(= 3 M_2) > m_{\tilde{q}} (=800$ GeV).  
Hence, $Br[\tilde{g} \to \tilde{q} q]=100 \%$. 
Effectively, then,  all electroweak cascades start with the squarks. 
Setting the higgsino mass parameter $\mu=150$ GeV makes the lighter 
``ino''s higgsino-like 
and degenerate thus closing the little cascades. 
Also, squarks of lighter families mainly decay to gaugino-like heavier 
chargino and neutralinos which undergo big cascades to produce Higgs bosons.
We try to probe the lacuna in the $M_A-\tan\beta$ space and so set $M_A=150$ GeV 
and $\tan\beta=5$.
This results in $M_h \simeq 110$ GeV (still not SM-like; hence LEP bound
does not apply) and $M_{H^\pm} \simeq 170$ GeV.
%($t \to H^+b$ for the canonical production mode would be strongly suppressed).

\begin{figure}[hbtp]
\vspace*{-2.70cm}
\hspace*{-1.9cm}
\epsfxsize=15.0cm   %width of figure - will enlarge/reduce the figures
\epsfbox{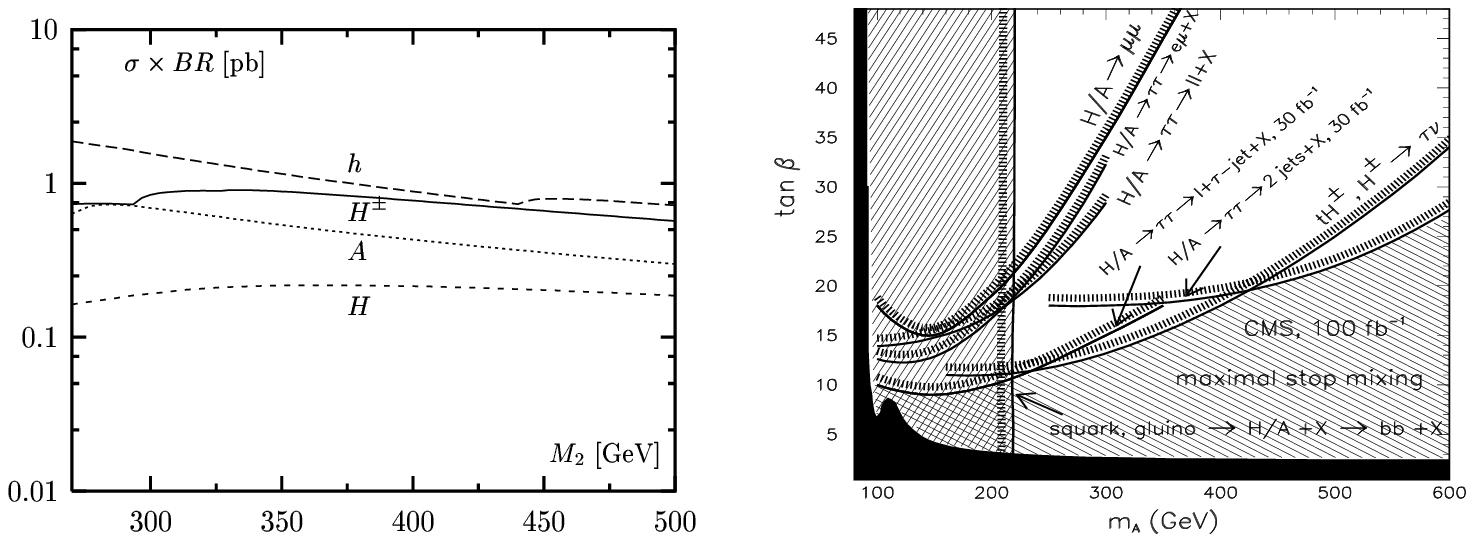}
\vspace*{-14.1cm}
\end{figure}
 
The figure on above left shows that rates for different MSSM Higgses lie between
0.1--1 pb for a range of $M_2$. Thus a large number of Higgs 
events (up to $\sim 10^4$ with $\int \mathcal L$=\,30 ${\mathrm fb}^{-1}$) is
expected in few years of LHC run.

Fast Monte Carlo simulation involves \cite{packages}
MSSM spectrum from ISASUSY v7.58
interfaced to HERWIG 6.4 by ISAWIG to generate the signal and the
backgrounds while detector response is included
through CMSJET 4.801. 

Large SUSY background comes from SUSY events not containing 
the Higgs bosons. Comparatively smaller SM one comes from the 
$t \bar t$ process. 
Basic kinematic distributions studied 
are the jet multiplicities, ${\not \!\! E}_T$  and $E_T^{jet}$. 

We then analyze the $b \bar b$ decay mode (BR $\sim 90 \%$) for the neutral
Higgses.
Suitable lower cuts are employed for all of the above variables. 
These help distinguish the reconstructed peaks for $h$ and $A,H$  
in the $b \bar b$ mass spectrum. 
For $H^\pm$ we choose the decay mode
$H^{\pm}\rightarrow \tau^{\pm} \nu_{\tau}$ (BR $\sim 95 \%$). We use similar cuts
but presence of neutrinos prohibits reconstruction of the mass peak. Hence, we further 
exploit the tau-polarization features \cite{dpsrc} with TAUOLA
\cite{packages} to tame the dominant
SM background from $W^{\pm}\rightarrow \tau^{\pm} \nu_{\tau}$. 
Even then, the evidence for $H^\pm$ is not as compelling as for neutral Higgses. 
However, this when combined with prior observation of  
neutral Higgses could bring forth a solid circumstantial evidence for $H^\pm$.

The figure on right summarizes reach for $\int \mathcal L$=\,100 ${\mathrm fb}^{-1}$
with $M_2$=350 GeV in this scenario.
In the hatched vertical column on left, heavier CP-even $H$ and pseudoscalar $A$ can be 
observed in the (big) cascades for $M_A \lesssim 220$ GeV for all $\tan\beta$. The corresponding
reach in $M_H^{\pm}$  is $\sim 200$ GeV. We see that this fills up the hole in the low
$m_A$ and intermediate $\tan\beta$ region and thus 
becomes \emph{complementary to the standard searches}.

Of course, there are generic scenarios where both little and 
big cascades may be present simultaneously \cite{H+cascade} which could
enhance the signal. We find, under favorable situation, 
one or more Higgs bosons with $M_\phi \lesssim$ 200 GeV can be probed
in SUSY cascades even with $\int \mathcal L$=\,30 ${\mathrm fb}^{-1}$.
Here, I do not discuss the cascades in eqs. (3) and (4) which would further 
enhance the yield of MSSM Higgses. Also, a better understanding of 
the SUSY backgrounds would help improve the overall reach.

\section*{Acknowledgments}
I wish to thank my collaborators A. Djouadi, M. Guchait and F. Moortgat.
I would also thank the organizers of SUSY 2003 for the nice meeting.
This work is supported by a DOE grant DE-FG02--97ER41029.

%%%%%%%%%%%%%%%%%%%%%%%%%%%%%%%%%%%%%%%%%%%%%%%%%%%%%%%%%%%%%%%%%%%%%%%

\end{document}